# Machine learning approaches for automatic cleaning of investigative drilling data

Fei Huang[1], Hongyu Qin[1, *], Masoud Manafi[2], Ben Juett[2], Ben Evans[2]

[1]Flinders University, Adelaide, Australia

[2]Civil Group (Aust) Pty Ltd, Adelaide, Australia

**Abstract**

Investigative drilling (ID) is an innovative measurement while drilling (MWD) technique that has been implemented in various site investigation projects across Australia. While the automated drilling feature of ID substantially reduces noise within drilling data streams, data cleaning remains essential for removing anomalies to enable accurate strata classification and prediction of soil and rock properties. This study employed three machine learning algorithms—IsoForest, one-class SVM, and DBSCAN—to automate the data cleaning process for ID data in rock drilling scenarios. Two data cleaning contexts were examined: (1) removing anomalies in rock drilling data, and (2) removing both anomalies and soil drilling data in mixed rock drilling data. The analysis revealed that all three machine learning algorithms outperformed traditional statistical methods (the 3σ rule and IQR method) in both data cleaning tasks, achieving a good balance between true positive rate and false positive rate, though hyperparameter tunings were required for one-class SVM and DBSCAN. Among them, IsoForest was proven to be the best-performing algorithm, capable of removing anomalies effectively without the need for hyperparameter adjustment. Furthermore, IsoForest, combined with two-cluster K-means, successfully eliminated both soil drilling data and anomalies while preserving almost all the normal data. The automatic data cleaning strategy proposed in this paper has the potential to reduce laborious manual data cleaning efforts and thereby facilitate the development of large-scale, high-quality datasets for machine learning studies capable of revealing complex relationships between drilling data and rock properties.

**Keywords**: investigative drilling, measurement while drilling, machine learning, IsoForest, one-class SVM, DBSCAN

## 1. Introduction

Investigative drilling (ID) is an emerging variant of measurement while drilling (MWD) technique that has been applied in various site investigation projects in Australia (Duthy & Juett, 2020; Huang et al., 2025). Ideally, the drilling data collected by MWD represents the response of geomaterials to drill bits and, thereby, offers valuable insights into their properties. Researchers have made considerable efforts to establish the relationships between drilling data and soil and rock properties through analytical analysis and laboratory or field drilling tests (Gui et al., 2002; Kalantari et al., 2018; Li & Itakura, 2012; Nishi et al., 1998; Rodgers et al., 2020). In recent years, machine learning has also been adopted to interpret MWD data and outperform classical methods in some cases (Fernández et al., 2023; Gao et al., 2024; García et al., 2022). However, due to the complex indentation, cutting and impact mechanisms in the drilling process (Chen & Labuz, 2006; Cheng et al., 2018; Jaime et al., 2015), the drilling data collected by MWD are often inherently noisy (Chen & Yue, 2015; Gui et al., 2002; Reiffsteck et al., 2018). Besides, vibrations of the drill rods (on-the-top sensors) and contaminated mud pulse signals (down-the-hole sensors) introduce extra noise to the drilling data (Rodgers et al., 2018; Tu et al., 2012). Finally, operational actions such as adding drill rods, pausing for sampling or in-situ testing, and borehole cleaning further complicate the drilling data (Yue et al., 2004). These noises present challenges for the interpretation of MWD data and hinder the large-scale industrial application of these developed models (Leung & Sheding, 2015). While the automatic drilling feature of ID can significantly reduce noise in the drilling data (Huang et al., 2025), data cleaning remains necessary to remove non-informative entries or anomalies caused by operational actions (see Section 2), thereby creating high-quality datasets for analysis.

To address the noise in MWD data, various filtering and smoothing techniques have been employed. Gui et al. (2002) used a mean filter to smooth the highly noisy MWD data. Reiffsteck et al. (2018) applied the moving average method to filter MWD data containing high-frequency noise. Hansen et al. (2024) utilized root mean square filtered MWD data for machine learning analysis. Other signal processing techniques, such as the continuous wavelet transformation and the Savitzky-Golay filter, have also been explored for filtering drilling signals (Lee & Lee, 2023; Silversides & Melkumyan, 2021). While these techniques effectively denoise data, they do not eliminate noise but instead smooth it through mathematical operations. This substitution of noise with interpolated values often distorts normal data to some extent. In cases where drilling data is abundantly available, removing noise rather than smoothing it is preferable to preserve the information in the normal data. Furthermore, traditional smoothing methods are better suited for periodic, high-frequency noise and may not effectively handle the occasional, non-periodic anomalies characteristic of ID data, as shown in Figure 1. Statistical methods such as the 3σ rule and the box-plot method have also been employed to remove unrealistic values in multi-feature MWD datasets (Navarro et al., 2021; van Eldert et al., 2020; Wang et al., 2021). However, they are originally designed for one-dimensional data, and comprehensive evaluations of their effectiveness in dealing with high-dimensional drilling data are limited. Finally, although manual checking is viable in removing anomalies in a small number of boreholes (Silversides & Melkumyan, 2021), it could be a daunting task when dealing with data from hundreds of boreholes. Machine learning, widely recognized for its ability to handle high-dimensional and large-scale data, offers a promising alternative for anomaly detection on big datasets (Nassif et al., 2021). Yet, its application in cleaning MWD or ID data remains largely unexplored. To advance automatic data cleaning in ID, this paper investigates the effectiveness of machine learning algorithms, including the isolation forest (IsoForest), one-class support vector machine (one-class SVM), and density-based spatial clustering of applications with noise (DBSCAN), in detecting and removing anomalies in rock drilling data. These methods are evaluated against traditional statistical techniques, such as the 3σ rule and interquartile range (IQR) method, to assess their relative performance in anomaly detection and data cleaning.

## 2. ID data streams

The ID rig normally collects drilling data every 5 cm (or smaller if required), producing continuous data streams along the depth. The abundant drilling data produced by ID enables the adoption of machine learning in the analysis, which could potentially unravel complex patterns behind the drilling data. Although noise has been significantly reduced in the drilling data of ID due to its automatic drilling feature (Huang et al., 2025), non-informative data or anomalies still exist in data streams. These data do not reflect the actual properties of geomaterials and could potentially deteriorate the performance of machine learning algorithms. Based on the analysis of the drilling process, there are four main drilling actions that cause anomalies in ID data streams, as illustrated in Figure 1.

- Collaring: In this stage, the drill bit initially penetrates the ground, and the drilling process has not reached a stable state. Data in this stage features increasing percussion pressure from zero to the designed value and very high penetration rate values.
- Unexpected pauses: Drilling is sometimes unexpectedly stopped, and the percussion pressure suddenly drops to a very low value, while the feed and rotation pressure first drop and then quickly jump up to a high value. The penetration rate is also very low due to the prolonged drilling time. Unexpected pauses do not always happen.
- Rod adding: The data response is similar to that in unexpected pauses. Rod adding occurs at intervals of one-rod length during drilling, which is 8 m for the drilling data in this paper.
- End of drilling: The drilling rig quickly reduces the percussion pressure and feed pressure, leading to a low penetration rate. Since the bit is still advancing, one or two data points are still recorded in this stage.

Although anomalies can be identified and removed through manual inspection based on the features described above, variations in their depth and length across boreholes make manual cleaning impractical for datasets with hundreds of boreholes.

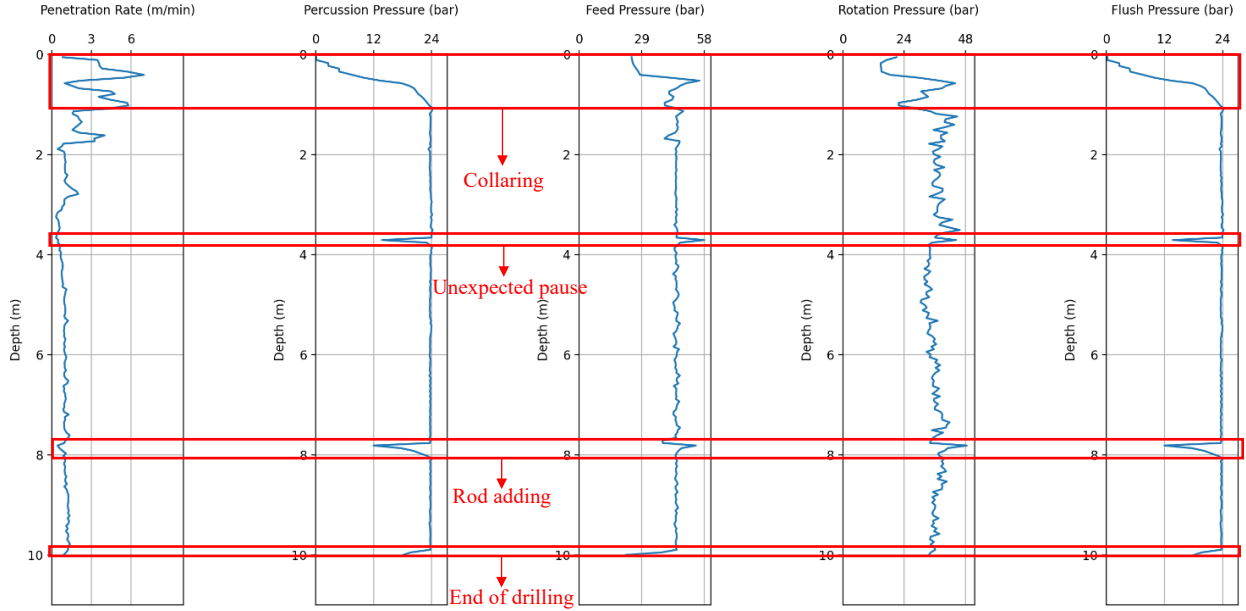

Figure 1 Anomalies in the ID data streams in percussive drilling of a typical borehole

Typical drilling parameters collected by the ID rig in percussive drilling are the penetration rate (m/min), percussion pressure (bar), feed pressure (bar), rotation pressure (bar) and flush pressure (bar). As illustrated in Figure 1, the percussion pressure and flush pressure are the same in percussive drilling. Therefore, the flush pressure was omitted in the investigation of different data cleaning methods. ID data were preserved following the standard data format in machine learning: every drilling data is a row with depth and drilling parameters in different columns. In this study, two data cleaning tasks are examined using ID data from two example boreholes: one with rock drilling data and anomalies (borehole A) and the other with an extra section of soil drilling data in addition to rock drilling data and anomalies (borehole B). The statistical characteristics of the drilling data of these two boreholes are summarized in Table 1.

Table 1 Statistical characteristics of the drilling data of the two example boreholes

|  | Penetration rate (m/min) | | Percussion pressure (bar) | | Feed pressure (bar) | | Rotation pressure (bar) | |
| --- | --- | --- | --- | --- | --- | --- | --- | --- |
|  | Borehole A | Borehole B | Borehole A | Borehole B | Borehole A | Borehole B | Borehole A | Borehole B |
| Count | 190 | 183 | 190 | 183 | 190 | 183 | 190 | 183 |
| Mean | 1.34 | 3.10 | 22.51 | 18.64 | 44.20 | 37.24 | 35.46 | 32.35 |
| Std. | 1.09 | 3.26 | 4.25 | 7.53 | 4.43 | 5.22 | 5.00 | 8.90 |
| Min. | 0.32 | 0.59 | 0.18 | 0.24 | 21.80 | 25.46 | 14.89 | 17.27 |
| 25% | 0.91 | 1.03 | 23.64 | 7.76 | 44.66 | 30.98 | 34.07 | 22.12 |
| 50% | 1.00 | 1.31 | 23.70 | 23.66 | 44.98 | 39.88 | 35.69 | 35.62 |
| 75% | 1.24 | 3.97 | 23.78 | 23.73 | 45.31 | 40.19 | 38.07 | 37.87 |
| Max. | 6.98 | 9.34 | 24.21 | 24.18 | 58.04 | 47.00 | 48.47 | 61.55 |

To facilitate evaluation using standard metrics such as the true positive rate (Recall) and false positive rate, anomalies from each of the identified drilling actions and soil responses were labeled as '-1', while normal data were labeled as '1' for the two example boreholes. A standard tool, the confusion matrix, was also employed to visualize these metrics. The distribution of anomalies coming from each source is detailed in Table 2.

Table 2 Detailed anomaly distribution for the two example boreholes

| | Collaring | Unexpected pauses | Rod adding | End of drilling | Transition from soil to rock | Soil drilling | Total anomalies | Total normal data |
|---|---|---|---|---|---|---|---|---|
| Borehole A | 19 | 1 | 6 | 2 | - | - | 28 | 162 |
| Borehole B | 7 | - | 8 | 2 | 9 | 39 | 65 | 118 |

## 3. Data cleaning methods

Three anomaly detection algorithms—IsoForest, one-class SVM, and DBSCAN—along with a clustering algorithm, K-means, were compared with two conventional statistical methods (the 3σ rule and the IQR method) in terms of their effectiveness in detecting and removing anomalies in ID data streams. The two statistical methods were implemented using self-compiled code in Python, while the four machine learning algorithms were implemented using the machine learning library, Scikit-learn (Pedregosa et al., 2011). It should be noted that the drilling data was scaled using StandardScaler, a feature scaling tool that transforms the features to have a mean of zero and a standard deviation of one, before being fed into one-class SVM, DBSCAN and K-means, which are sensitive to data scales.

*3.1 Conventional statistical methods*

*3.1.1 The 3σ rule*

The 3σ rule is a commonly used statistical method for anomaly detection and data cleaning. The rule assumes that the data follows a normal distribution and that the possibility of a data point falling within three standard deviations (3σ) of the mean is about 99.7%. The possibility of any data points that fall outside this range is so small that they are typically considered anomalies and should be removed from the dataset. To clean high-dimensional data like ID data in this study, the 3σ rule is applied separately to each feature. If any value in a row is flagged as an anomaly in any feature, the entire row is removed from the dataset, even if the remaining parameters in that row are within normal ranges.

*3.1.2 The IQR method*

The IQR method is another widely used statistical method in data cleaning to remove outliers or anomalies. It is based on the concept of dividing data into quartiles (the 25th percentile and 75th percentile) and measuring the spread of the middle 50% of the data. The *IQR* is calculated as:

$$IQR = Q_3 - Q_1 \qquad (1)$$

where $Q_1$ is the 25th percentile of the data, and $Q_3$ is the 75th percentile of the data. Anomalies are data points that fall outside the range defined by the lower bound, $Q_1-1.5 \times IQR$, and the upper bound, $Q_3-1.5 \times IQR$.

When using the IQR method for cleaning ID data, the same procedure as the 3σ rule is adopted: remove the entire row when the data in a feature is detected as an anomaly. The IQR method is more suitable for detecting outliers in skewed data compared to the 3σ rule.

*3.2 Machine learning algorithms*

*3.2.1 IsoForest*

IsoForest is an unsupervised decision-tree-based machine learning algorithm for anomaly detection (Liu et al., 2008, 2012). It relies on the assumption that anomalies are rare and distinct from the rest of the data, allowing them to be isolated with a limited number of splits in a binary tree. IsoForest works by randomly selecting a feature from the dataset and splitting the dataset by a randomly selected split point of that feature. The partition process is repeated until the maximum tree height is reached or there is only one data point (or only the same points) in the leaf node. The path length from the leaf node to the root node, *h(x)*, is used to assess whether a data point is an outlier or not. For a collection of isolation trees in the IsoForest, the average path length of a data point is *E(h(x))*, then the anomaly score of a data point *s(x)* is defined as:

$$s(x,n) = 2^{-\frac{E(h(x))}{c(n)}} \tag{2}$$

where *n* is the sample number in the dataset; *c(n)* is the approximation of the average path length, and can be expressed as:

$$c(n) = 2H(n-1) - (2(n-1)/n) \tag{3}$$

where *H* is the harmonic number approximated by $\ln(i) + 0.57721$.

If the anomaly score of a data point is close to 1, then this data point will be regarded as an anomaly. IsoForest is efficient and capable of handling large and high-dimensional datasets.

*3.2.2 One-class SVM*

One-class SVM is an unsupervised machine learning method for novelty or anomaly detection (Schölkopf et al., 1999). It is a variant of the traditional support vector machine algorithm, which is usually used in supervised machine learning for classification and regression (Vapnik, 1997). For anomaly detection, one-class SVM constructs a boundary (hyperplane) around the training data in the feature space to separate them from the origin. The algorithm then tries to maximize the margin between the boundary and the origin. For data points located at the origin side, the algorithm will identify them as anomalies. To maximize the margin, one-class SVM tries to solve the following optimization problem:

$$\min_{w,\rho} \frac{1}{2}\|w\|^2 + \frac{1}{\upsilon n}\sum_{i=1}^{n}\xi_i - \rho \tag{4}$$

subject to:

$$w \cdot \phi(x_i) \geq \rho - \xi_i, \ \xi_i \geq 0 \tag{5}$$

where $\phi$ is the mapping of data points into a higher-dimensional space (using a kernel); *v* is the parameter that controls the proportion of outliers; $\rho$ is the offset that defines the boundary; $\xi$ is the slackness variable that allows some data points to fall outside the boundary (to account for outliers).

One-class SVM can identify non-linear decision boundaries when using kernel tricks, but can also be computationally expensive when the dataset is large.

3.2.3 DBSCAN

DBSCAN is a popular unsupervised clustering algorithm that identifies clusters based on the density of data points (Ester et al., 1996). In the concept of DBSCAN, if there are at least *MinPts* within a radius of $\varepsilon$ of a point *p*, then it is called a core point; a point *q* is called *directly reachable* from *p* if it is within $\varepsilon$ of *p*, or *reachable* from *p* if it is not within $\varepsilon$ of *p* but can be reached through a path by *p* where each point is directly reachable from its neighbors. The core point and all other points that are directly reachable or reachable from it form a cluster. DBSCAN starts visiting an arbitrary point that has not been visited. If that point is a core point, the algorithm marks it and its neighbors as part of the same cluster. If it is not a core point, the algorithm marks it as noise but can be revisited if it later becomes part of a cluster. DBSCAN recursively visits all neighbors of the core point and its neighbors to expand the cluster until no more points meet the criteria. In anomaly detection, DBSCAN classifies the data points that do not belong to any cluster as anomalies. Although DBSCAN does not need a predefined cluster number and holds no assumption on the cluster shape, the two hyperparameters, $\varepsilon$ and *MinPts*, should be carefully chosen to produce satisfactory anomaly detection results.

*3.2.4 K-means*

K-means is another popular clustering algorithm in unsupervised machine learning (Lloyd, 1982; MacQueen, 1967). K-means aims to minimize the within-cluster sum of squares (*WCSS*):

$$WCSS = \sum_{i=1}^{k} \sum_{x \in C_i} \|x - \mu_i\|^2 \tag{6}$$

where $C_i$ is the set of data points in cluster $i$; $\mu_i$ is the centroid of cluster $i$; $\|x - \mu_i\|$ is the distance between a data point $x$ and the centroid $\mu_i$; $k$ is the predefined cluster number.

The commonly used distance metric is Euclidean distance which is defined by:

$$d(x_i, x_j) = \sqrt{\sum_{k=1}^{n} (x_{ik} - x_{jk})^2} \tag{7}$$

where $x_i$ and $x_j$ are two data points with $n$ dimensions or features, and $k$ is the index of the dimensions.

K-means achieves its objective (minimizing *WCSS*) through an iterative refinement process. Initially, the algorithm selects $k$ random centroids. It then calculates the distance between each data point and the centroids in the feature space. Based on these distances, each data point is assigned to the cluster with the nearest centroid. The centroids are then updated by calculating the mean of the data points in each cluster. This cycle of distance calculation, cluster assignment, and centroid update continues until convergence, which occurs when the centroids no longer change. K-means is a straightforward and easy-to-implement clustering algorithm. However, it is not inherently designed for anomaly detection. In this study, K-means is used to assist Isoforest in further separating the soil drilling data from cleaned rock drilling data.

## 4. Results and discussion

*4.1 Data cleaning for rock drilling data with anomalies*

The first data cleaning task examined in this paper is the removing of anomalies from rock drilling data. A borehole (borehole A) with 190 data points was used in the analysis. The data cleaning results using the 3σ rule are illustrated in Figure 2. Overall, 13 out of the 28 total anomalies were detected. Specifically, 11 out of the 19 anomalies in the collaring stage were detected, while the single anomaly caused by the unexpected pause at 3.7 m was successfully identified. For the 6 anomalies caused by rod adding, none of them were detected by the 3σ rule, whereas one of the two anomalies at the end of drilling was identified. Upon further examination, the data cleaning effectiveness of the 3σ rule mainly comes from the penetration rate and feed pressure, which eliminate the overly high penetration rate in the collaring stage and overly low feed pressure in the collaring and end-off-drilling stages, respectively. It can be concluded that the 3σ rule missed many anomalies and is not very effective for this task.

Figure 3 presents the data cleaning results using the IQR method. While it successfully detected all 28 anomalies, it severely overestimated the number of outliers by incorrectly removing 34 normal data points, particularly those occurring immediately after the collaring stage and between depths of 3 to 8 meters. This over-cleaning is largely due to the highly uniform percussion and feed pressures during normal drilling, which led the IQR method to define very narrow acceptable ranges for these parameters (23.44–23.98 bar and 43.69–46.28 bar, respectively). As a result, the method not only removed anomalies but also filtered out normal variations that reflect changes in rock properties. These findings suggest that the IQR method is too stringent for effectively cleaning anomalies in ID rock drilling data.

Since the 3σ rule and IQR method are primarily designed for one-dimensional data, Principal Component Analysis (PCA) was applied to reduce the four-dimensional drilling dataset to a single dimension (the first principal component). This transformation aimed to enable a fairer comparison with machine learning algorithms, which can handle high-dimensional data directly. The cleaning results of the 3σ rule and IQR method, both with and without PCA, are illustrated in Figure 4 using the percussion pressure as a representative parameter. The results show that after dimensionality reduction, both methods identified fewer anomalies: the number of detected anomalies decreased from 13 to 7 for the 3σ rule and from 62 to 20 for the IQR method, respectively. For the 3σ rule, PCA further compromised its anomaly detection performance. In contrast, PCA mitigated the over-strictness of the IQR method, reducing false positives to just three misclassified normal data points. However, the number of anomalies successfully identified by PCA+IQR remains relatively low compared to the machine learning algorithms introduced in the following section. Considering the extra efforts and limited effectiveness, PCA was not employed in the second data cleaning task.

Figure 5 shows the data cleaning results of IsoForest. It can be seen that IsoForest effectively cleaned 18 out of the 19 anomalies in the collaring stage, as well as the single anomaly produced by the unexpected pause and the two anomalies at the end of drilling. For the 6 anomalies caused by rod adding, IsoForest only removed the two data points with the most dramatic percussion pressure drops. While successfully cleaning most of the anomalies (23 out of 28), IsoForest greatly maintained the normal data, with only 4 out of 162 misidentified as anomalies. It should be noted that IsoForest was executed with the auto mode, which means no effort is needed to select the optimal hyperparameter for the best cleaning performance. This is a superior advantage of IsoForest for larger-scale data cleaning and will be shown more clearly in the next data cleaning task, where one-class SVM and DBSCAN need different optimal hyperparameters.

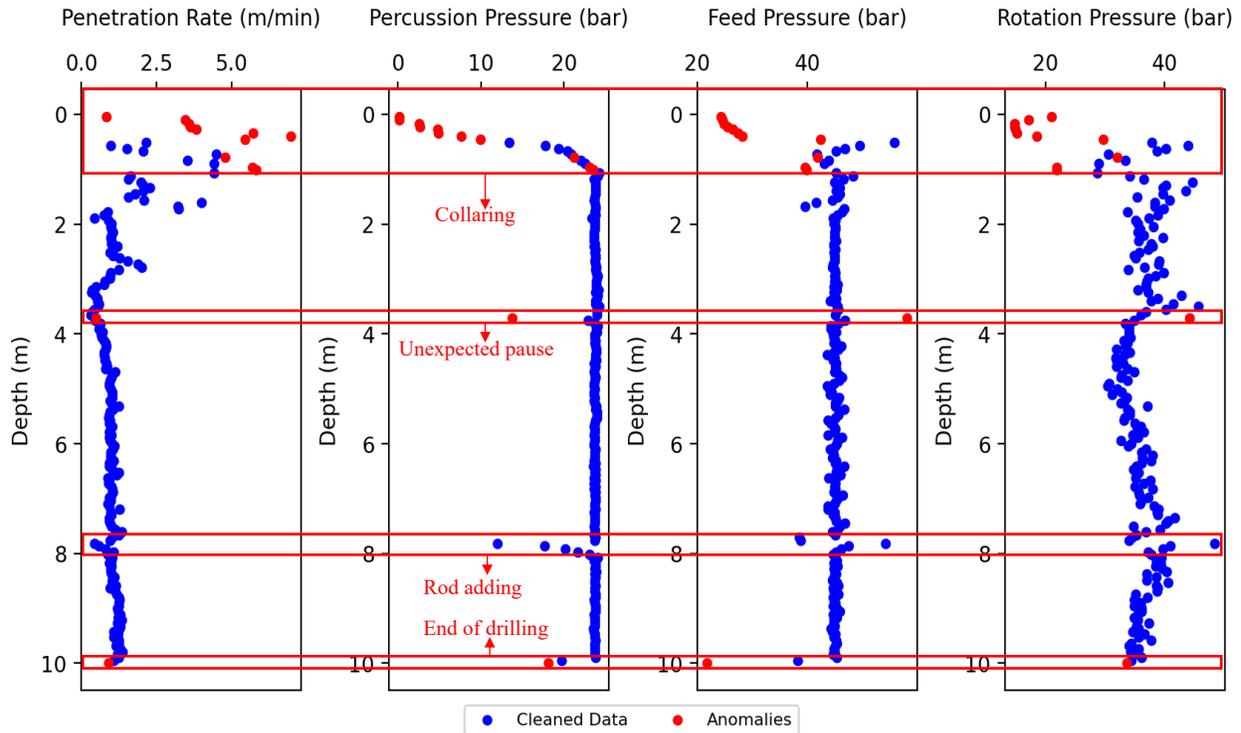

Figure 2 Cleaning results of the rock drilling data using the 3σ rule

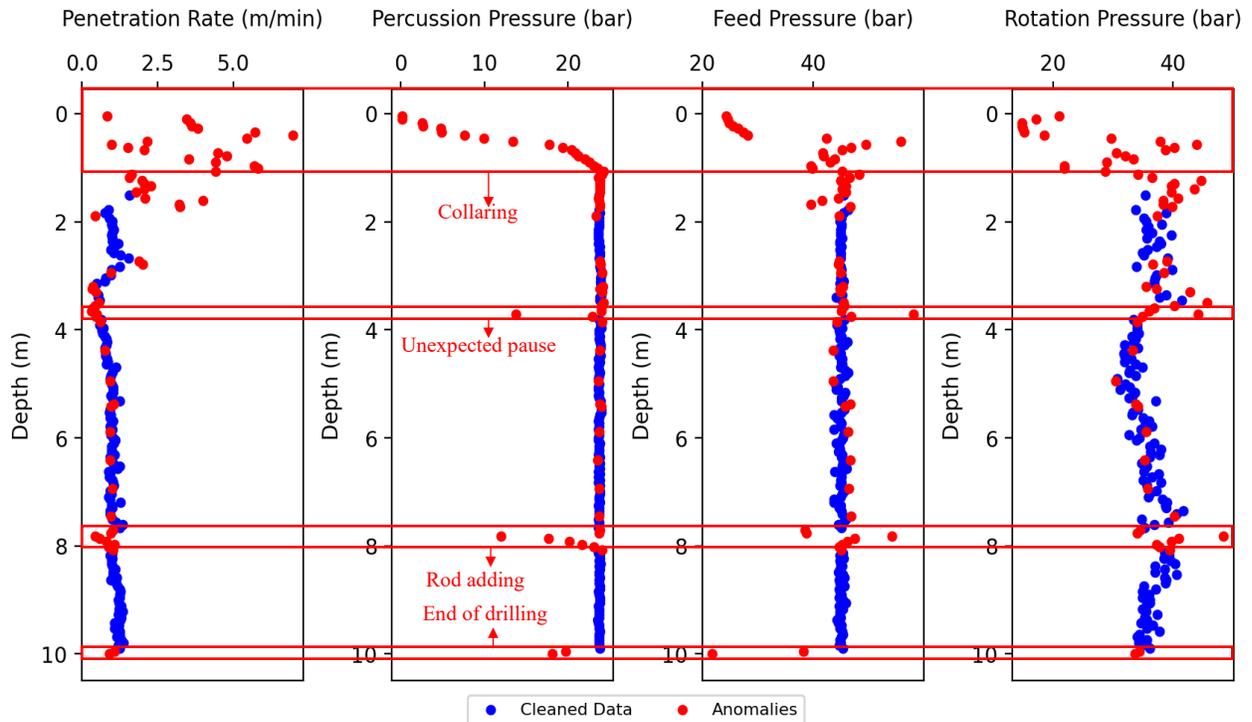

Figure 3 Cleaning results of the rock drilling data using the IQR method

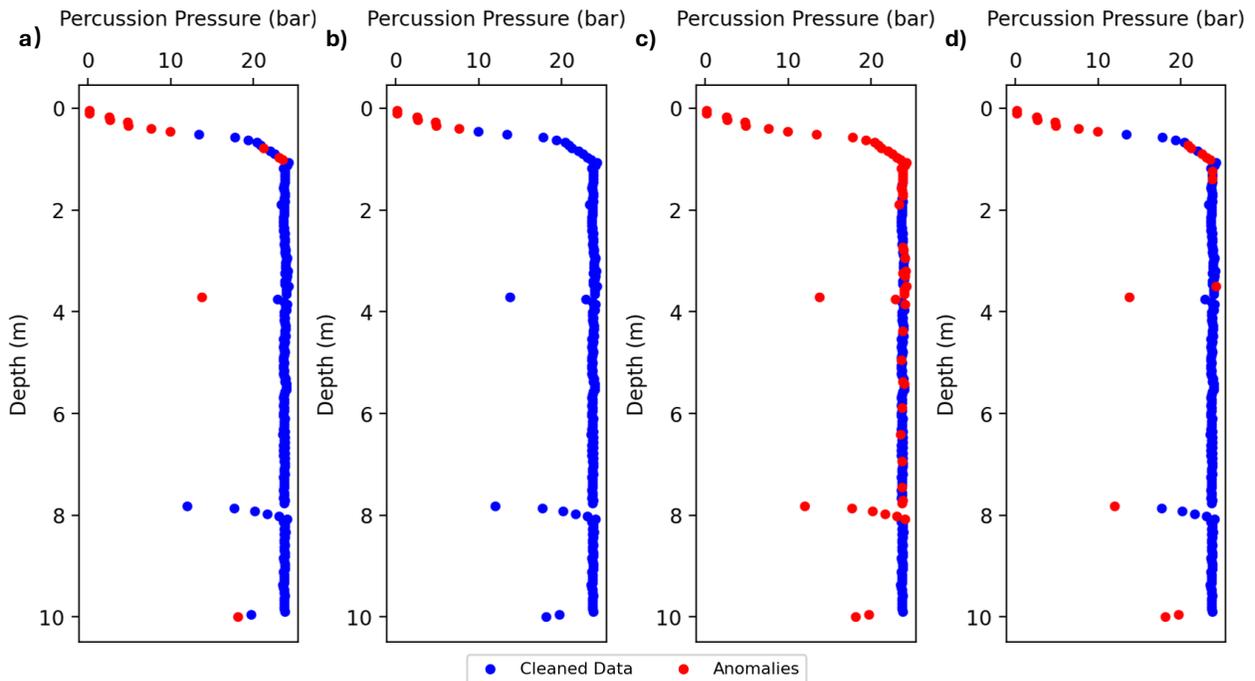

Figure 4 Cleaning results of the 3σ rule and IQR method with and without PCA: a) 3σ rule without PCA, b) 3σ rule with PCA, c) IQR method without PCA, d) IQR method with PCA

One-class SVM was also applied to clean the same rock drilling data. The performance of one-class SVM depends on the hyperparameter *nu*, which controls the proportion of data classified as anomalies—higher *nu* values lead to more aggressive anomaly detection. To ensure a fair comparison with IsoForest, *nu* was tuned through a trial-and-error process to match IsoForest's cleaning performance in the rod-adding section, removing the two data points with the most significant drops in percussion pressure. Figure 6 illustrates the cleaned data represented by percussion pressure and the detected anomalies with *nu* values between 0.1~0.3. As we can see, the algorithm is sensitive to the change of *nu*: the number of detected anomalies increases steadily with increasing *nu*. Among them, *nu* = 0.2 matches the aforementioned baseline. Figure 7 presents the cleaning results of one-class SVM with *nu* = 0.2, showing that the method successfully detected 25 anomalies across collaring, unexpected pauses, rod-adding, and end-of-drilling stages. Notably, it identified two more anomalies in the rod-adding stage than IsoForest. However, one-class SVM also misclassified 14 of the 162 normal drilling data points as anomalies, which is significantly higher than IsoForest. This could cause a considerable information loss for further analysis.

For DBSCAN, two governing hyperparameters, *eps* and *min_samples*, were tuned to the baseline performance of IsoForest detailed forehead. A trial-and-error procedure was adopted again for tuning the two hyperparameters of DBSCAN. After a few initial trials of different *eps* and *min_samples* values, it was found that the number of detected anomalies was more sensitive to *eps* than *min_samples*. Especially when *eps* reaches 0.4, *min_samples* had a minor influence on the detected anomaly numbers, and was thereby set to 10 since then. The tuning process of *eps* (0.4-0.8) is shown in Figure 8 with the variation of detected anomaly numbers. It can be seen that fewer data are identified as anomalies with the increase of *eps,* and the matching *eps* is found to be 0.7. Figure 9 illustrates the detailed cleaning results of DBSCAN with *eps*=0.7 (*min_samples*=10). It can be observed that DBSCAN effectively identified 26 out of the 28 total anomalies, with one more identified anomaly in the collaring stage than IsoForest and one-class SVM, and two more identified anomalies in the rod-adding stage than IsoForest. Additionally, DBSCAN misidentified 6 out of the 162 normal data points, which is slightly higher than IsoForest. From the perspective of cleaning performance, DBSCAN can be regarded as effective as IsoForest. However, more effort is needed to tune the two hyperparameters to achieve a good performance. The optimal hyperparameters could also be different for drilling data of different boreholes, which will significantly restrict its application in automatic data cleaning of large-scale datasets containing many boreholes.

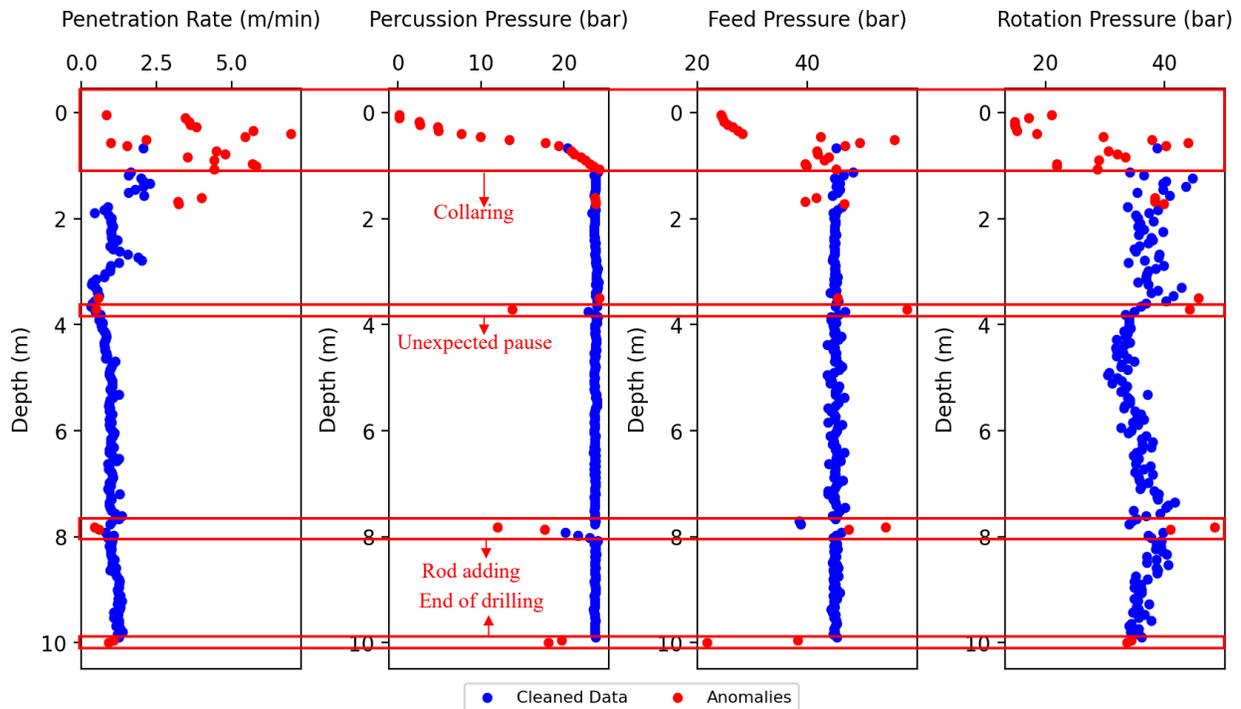

Figure 5 Cleaning results of the rock drilling data using IsoForest

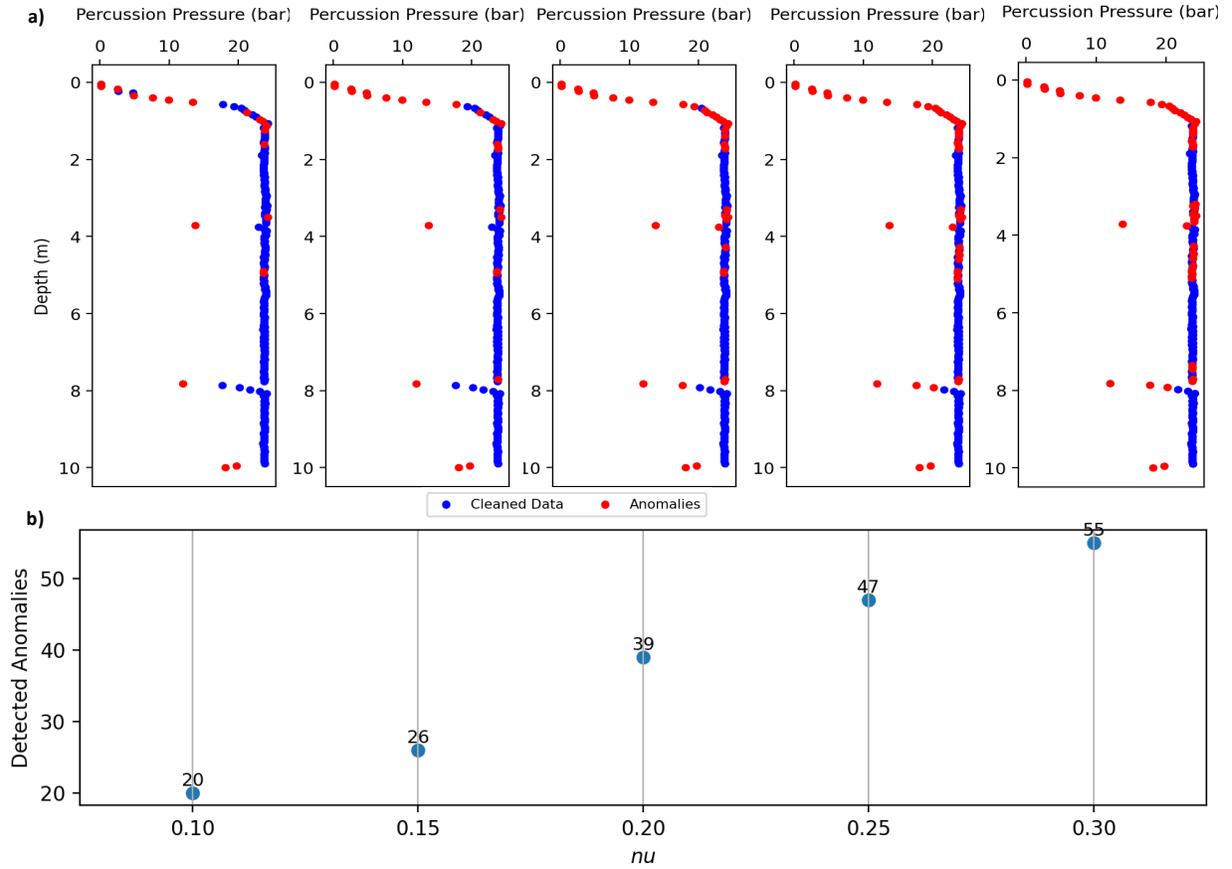

Figure 6 Tuning of *nu* for one-class SVM: a) cleaning results shown by the percussion pressure at different *nu* values, b) number of detected anomalies as a function of *nu*

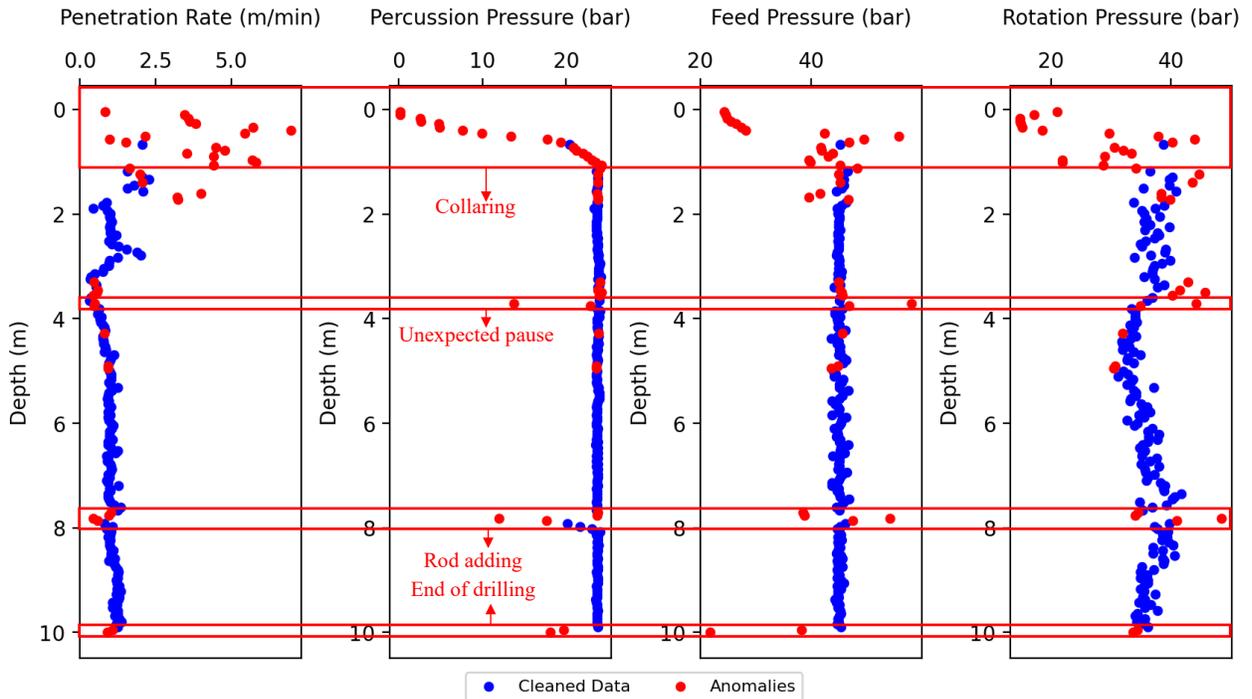

Figure 7 Cleaning results of the rock drilling data using one-class SVM (*nu*=0.2)

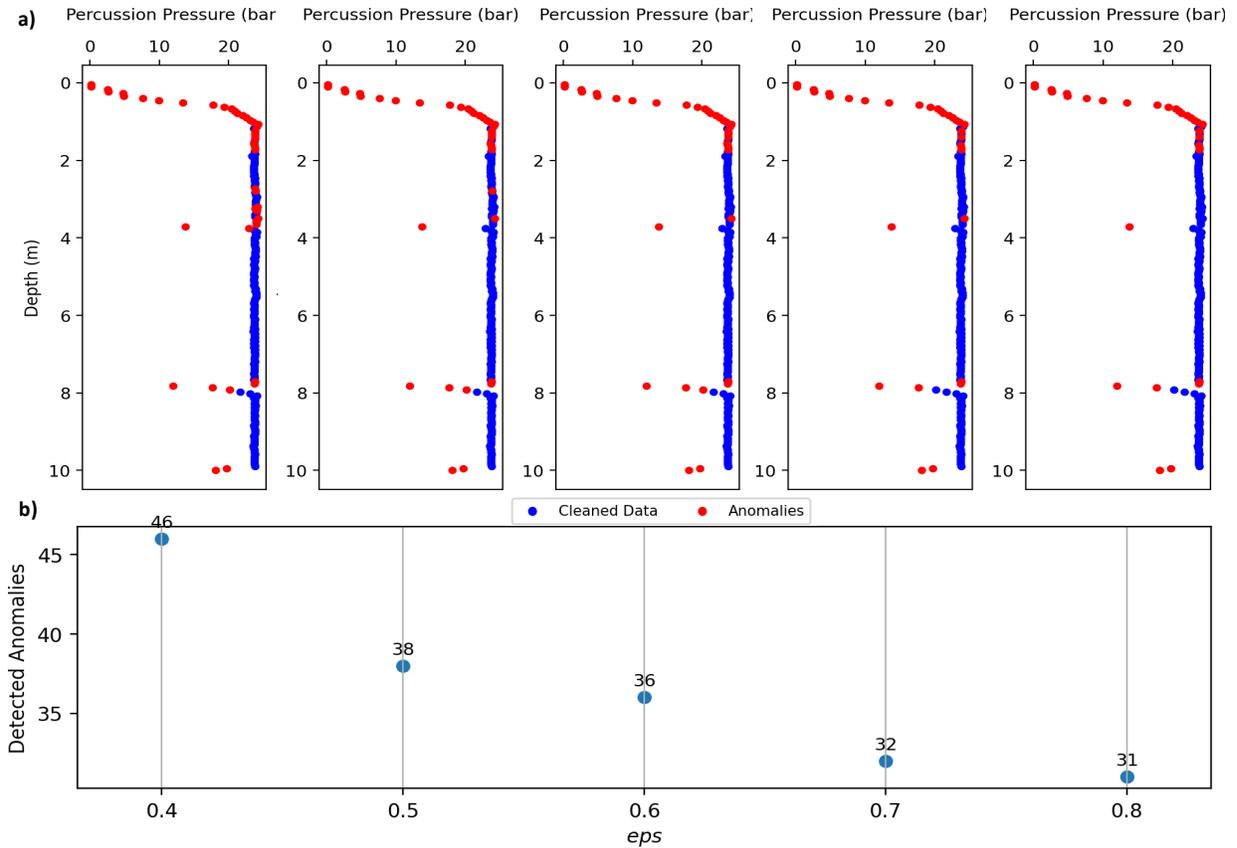

Figure 8 Tuning of *eps* for DBSCAN: a) cleaning results shown by the percussion pressure at different *eps* values, b) number of detected anomalies as a function of *eps*

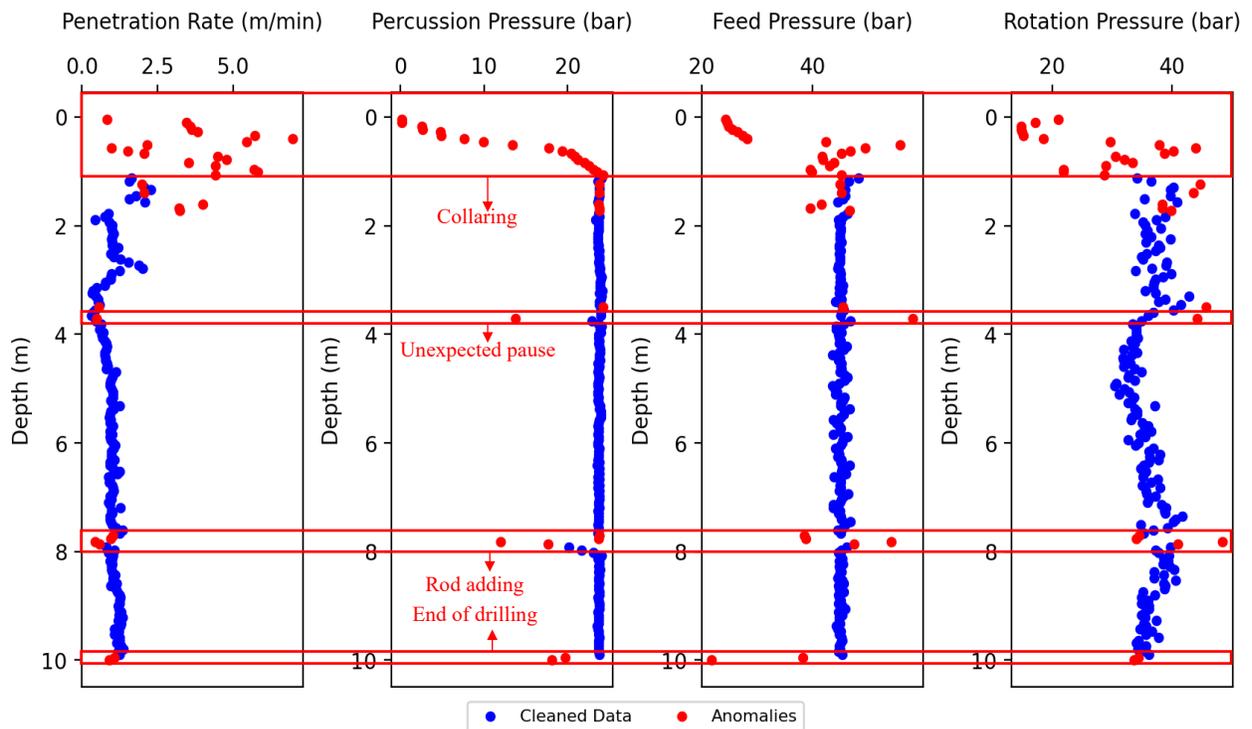

Figure 9 Cleaning results of the rock drilling data using DBSCAN (*eps*=0.7, *min_samples*=10)

Table 3 summarizes the detailed anomaly detection results of the 3σ rule, IQR method and three selected machine learning algorithms, as well as their average runtime of 6 repetitions. Figure 10 further illustrates these results using normalized confusion matrices, highlighting commonly used classification metrics such as true positive rate (Recall) and false positive rate. The true positive rate, shown in the bottom-right quadrant, represents the proportion of correctly identified anomalies, while the false positive rate, in the top-right quadrant, indicates the proportion of misclassified normal data as reported in Table 3.

Overall, all three machine learning algorithms outperformed the two statistical methods by achieving high true positive rates (0.82–0.93) and low false positive rates (0.025–0.086). In terms of computational costs, IsoForest has the longest runtime (0.077 s), while one-Class SVM and DBSCAN are significantly faster, requiring only 0.002 s and 0.003 s, respectively. However, the strong performance of one-Class SVM and DBSCAN depend on careful hyperparameter tuning, which cannot be automated across multi-borehole datasets due to the variability in optimal parameters between boreholes. As a result, the time spent on tuning would scale, offsetting their advantage in runtime. In contrast, IsoForest demonstrated a strong performance without requiring hyperparameter tuning, and its longer runtime is acceptable for large projects containing hundreds to thousands of boreholes due to its still relatively low single-borehole runtime ($10^{-2}$).

Regarding the robustness of IsoForest, it may be attributed to its underlying principle—anomalies are "few and different"—which leads to shorter path lengths for anomalies in the isolation trees (Liu et al., 2008). This behavior is intrinsic to the data distribution and doesn't rely heavily on distance metrics or density thresholds like one-class SVM and DBSCAN. Another reason for the hyperparameter insensitivity of IsoForest may be the random selection of features and thresholds to split nodes, which helps it generalize well across different data shapes and types. This hyperparameter insensitivity has also been reported for other tree-based machine learning algorithms, such as Random Forest and XGBoost (Probst et al., 2019; Wong & La, 2024).

Table 3 Summary of data cleaning results of the statistical methods and machine learning algorithms (borehole A)

| Data cleaning methods | Detected anomalies/total anomalies | Misidentified normal data/total normal data | Optimal hyperparameters | Average running time of 6 repetitions |
|---|---|---|---|---|
| 3σ rule | 13/28 | 0/162 | None | 0.003s |
| IQR method | 28/28 | 34/162 | None | 0.004s |
| IsoForest | 23/28 | 4/162 | Auto mode | 0.077s |
| One-class SVM | 25/28 | 14/162 | $nu$=0.2 | 0.002s |
| DBSCAN | 26/28 | 6/162 | $eps$=0.7, $min\_samples$=10 | 0.003s |

*Note:* All the data cleaning methods are performed on the CPU 13th Gen Intel(R) Core(TM) i5-1335U (1.30 GHz).

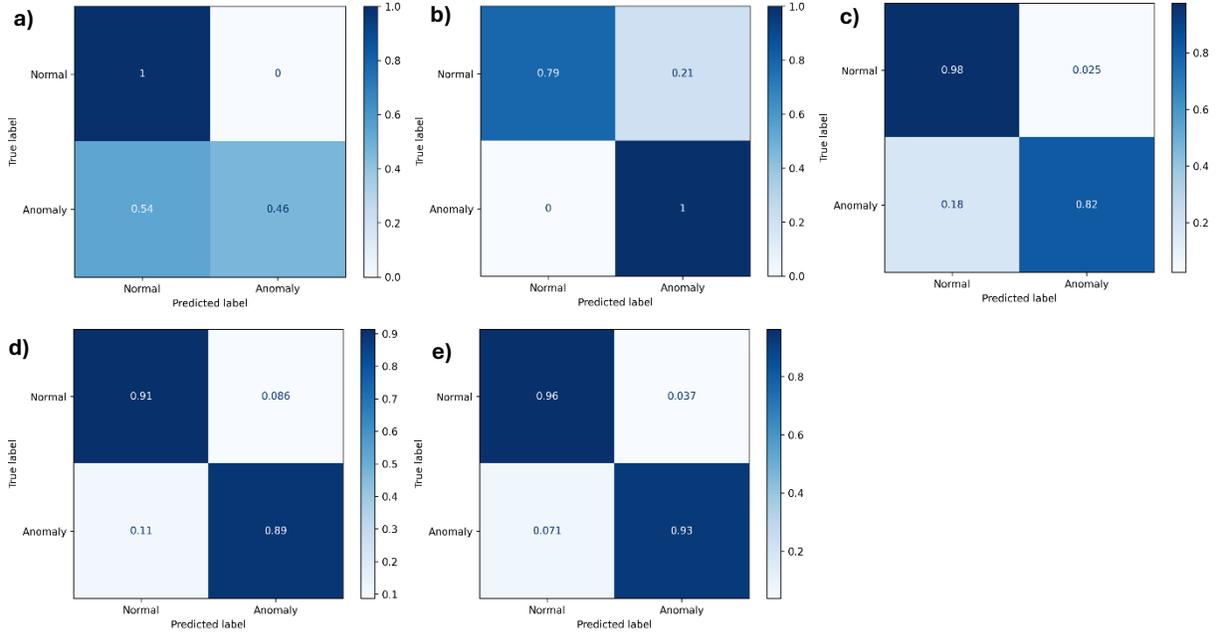

Figure 10 Normalized confusion matrices of the five data cleaning methods: a) 3σ rule, b) IQR method, c) IsoForest, d) one-class SVM, e) DBSCAN

*4.2 Data cleaning for rock and soil drilling data with anomalies*

A more complex data cleaning task involving the removal of both anomalies and soil drilling data from mixed rock drilling data is also investigated in this study. For this purpose, another borehole (borehole B) with 183 data points was selected. A length of 55 data points, containing sections of collaring, normal soil drilling and soil-rock transition, exists in this borehole above the rock drilling data. The soil-rock transition zone signifies the drill bit's exit from the soil and entry into the rock, resulting in unstable drilling responses. Data points within this zone were also treated as anomalies in the analysis.

First, the 3σ rule was employed to clean anomalies and soil drilling data. In Figure 11, it can be observed that only one anomaly in the soil-rock transition zone was detected. This is attributed to the cleaning effect on the rotation pressure column, which removed the data point with an exceptionally high rotation pressure. Since the existence of soil drilling data, the standard deviations of the drilling parameter are much larger than those in rocks only, which eventually disabled the anomaly detection capacity of the 3σ rule.

The IQR method was also applied to remove anomalies and soil drilling data from borehole B. Interestingly, as shown in Figure 12, it successfully eliminated all soil drilling data from the ID data streams, but only one anomaly in the transition zone was removed. Due to the large quantity of soil data and their distinctly different characteristics—such as very high penetration rates and very low percussion pressures—the resulting acceptable ranges became overly broad. As a result, only the high plateau in the penetration rate profile and the unusually high rotation pressure value fell outside these ranges. Overall, these findings indicate that the two statistical methods either failed to remove the anomalies or both the soil data and anomalies.

The data cleaning results of IsoForest for this task are presented in Figure 13. IsoForest successfully detected all 7 anomalies in the collaring stage, all two anomalies at the end of drilling, and most anomalies (8 out of 9) in the transition zone. It also identified 5 of the 8 most extreme anomalies caused by rod adding. Additionally, IsoForest demonstrated a low false positive rate, misidentifying only two normal data points. However, despite its effectiveness in anomaly detection, IsoForest struggled to remove soil drilling data, with only 5 out of 39 soil data points being eliminated. To address this limitation, further data processing techniques for removing soil drilling data will be explored in the following section.

Next, one-class SVM was employed to clean both anomalies and soil drilling data in borehole B. After tuning, the hyperparameter *nu* was found to be 0.3 to match IsoForest's performance in detecting anomalies in the rod-adding section. The tuning process of one-class SVM and DBSCAN for this task is the same as what was demonstrated in Section 4.1 and therefore not displayed in this section. As shown in Figure 14, one-class SVM identified most of the anomalies (6 out of 7) in the collaring stage and all anomalies in the soil-rock transition zone and at the end of drilling. It also removed nearly half of the 39 soil drilling data points. However, this method exhibited a higher false positive rate, misclassifying 16 normal data points as anomalies—an increase compared to the first data cleaning task.

Finally, DBSCAN was applied to this task, with its results presented in Figure 15. DBSCAN achieved nearly the same cleaning effectiveness and misidentification rate as IsoForest, only missing one anomaly in the soil collaring stage. Like IsoForest, DBSCAN was less effective in removing soil drilling data from the dataset. Significant effort was made to tune its hyperparameters, *eps* and *min_samples*, which were found to be 0.35 and 30 for this task, differing from the values (0.7 and 10) used in the first data cleaning task. This variation highlights the challenges of automating data cleaning across multiple boreholes. A summary of the detailed data cleaning results for DBSCAN and the other methods is presented in Table 4. It is evident that both statistical methods failed to detect anomalies in this more complex data cleaning task, although the IQR method successfully removed all soil drilling data. In contrast, all three machine learning algorithms identified most of the anomalies, with additional hyperparameter tuning required for one-class SVM and DBSCAN. However, without further intervention, none of the machine learning methods were able to effectively remove the soil drilling data.

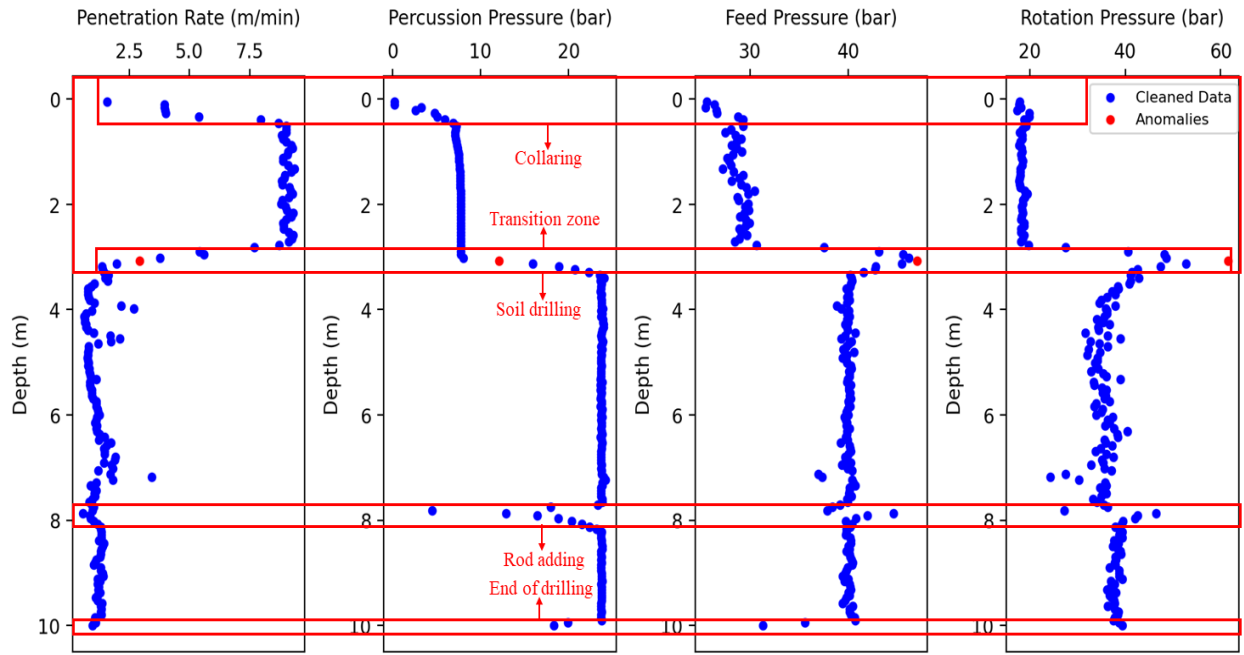

Figure 11 Cleaning rock and soil drilling data using the 3σ rule

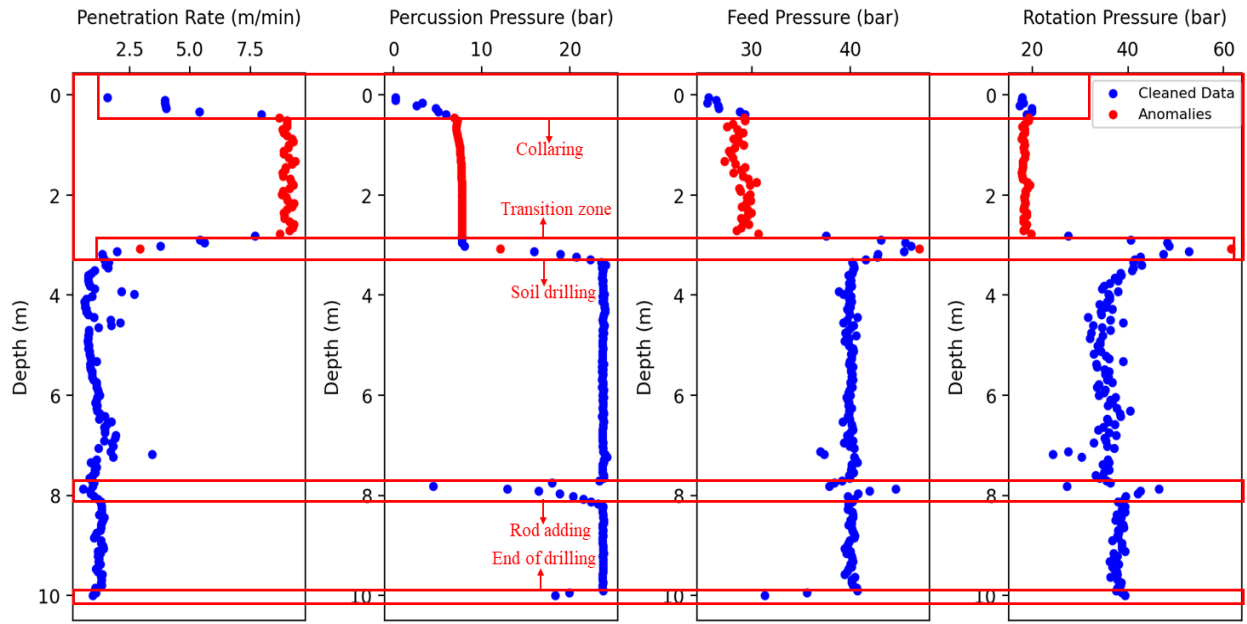

Figure 12 Cleaning rock and soil drilling data using the IQR method

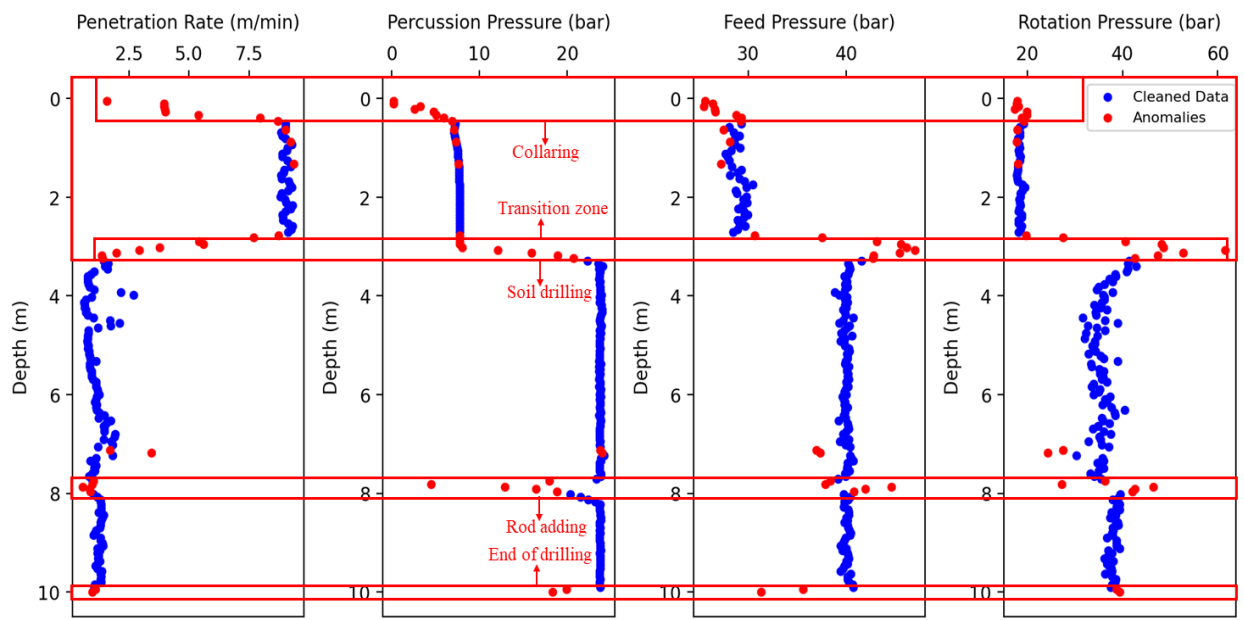

Figure 13 Cleaning rock and soil drilling data using IsoForest

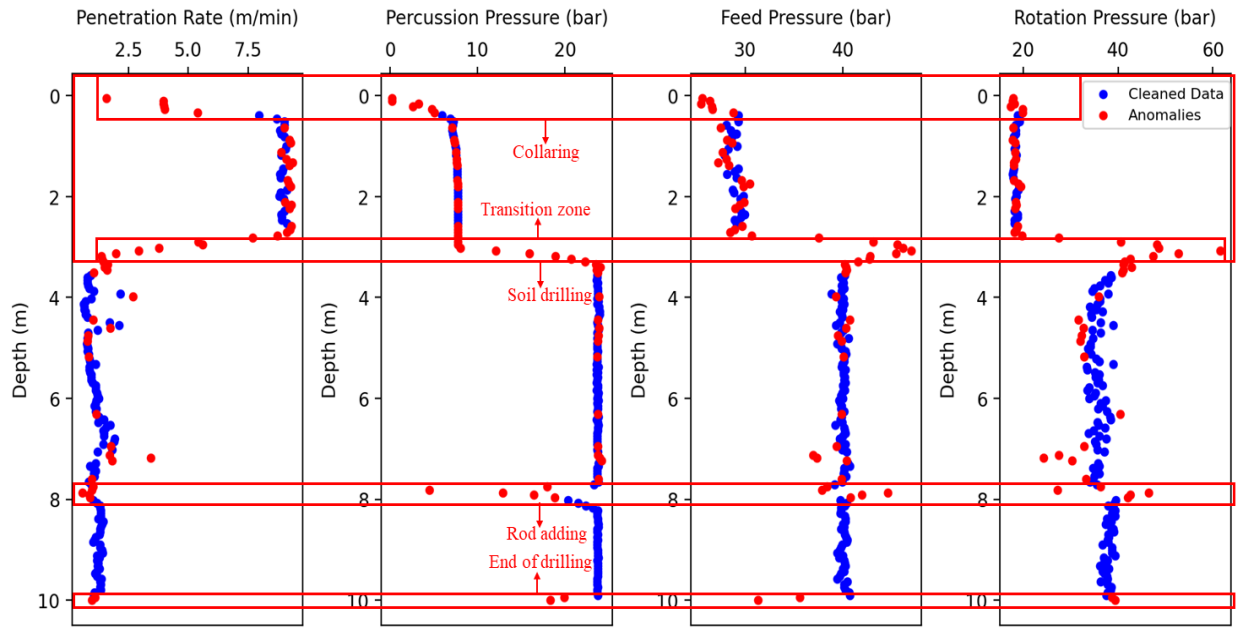

Figure 14 Cleaning rock and soil drilling data using one-class SVM (*nu*=0.3)

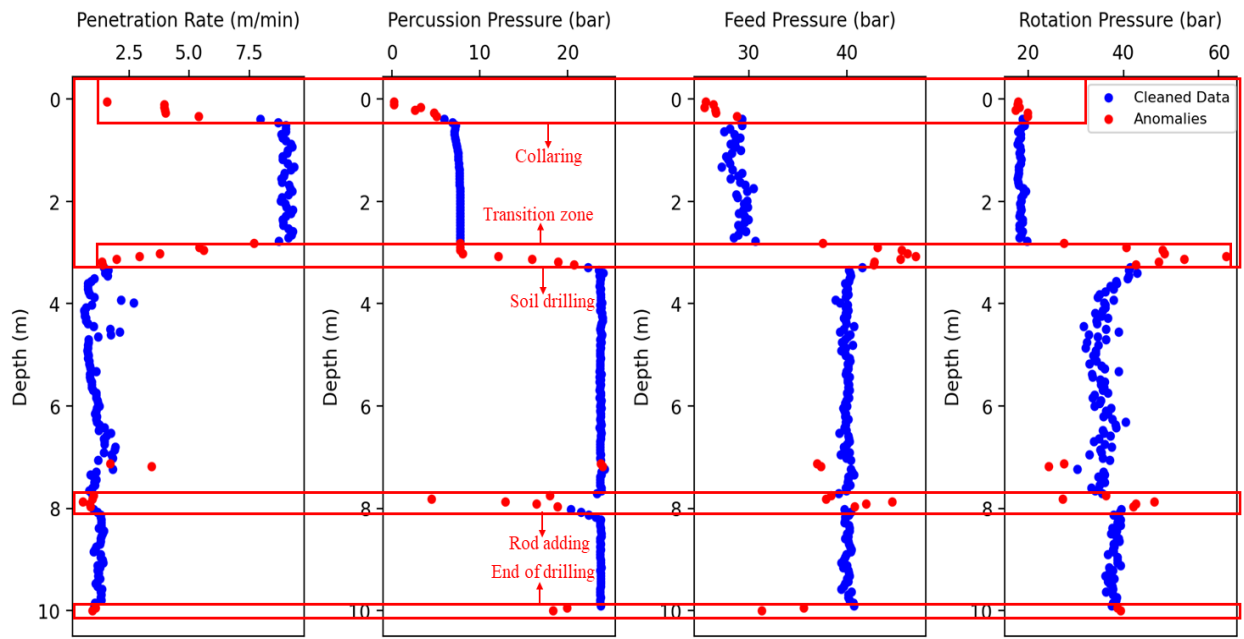

Figure 15 Cleaning rock and soil drilling data using DBSCAN (*eps*=0.35, *min_samples*=30)

Based on the findings from the two data cleaning tasks, it can be concluded that IsoForest demonstrates the best performance in detecting anomalies and preserving normal data across both tasks. However, it remained ineffective in removing soil drilling data. Given the distinct characteristics between soil and rock drilling data, a hybrid data cleaning strategy is proposed here to address the more complex challenge of eliminating both anomalies and soil data. This strategy involves two steps: first, IsoForest is applied to detect and remove anomalies in the ID data streams; second, K-means clustering with two clusters is employed to separate and remove soil drilling data from the IsoForest-cleaned dataset. The final cleaning results using this hybrid strategy are presented in Figure 16. It is evident that all remaining soil drilling data are effectively removed by the two-cluster K-means without discarding any rock drilling data. Combining the auto-mode IsoForest and two-cluster K-means, this approach enables automatic cleaning of rock drilling data containing both anomalies and soil interferences without the need for any manual hyperparameter setting.

In real-world projects, boreholes may consist of either entirely rock or mixed soil–rock profiles. To further examine whether applying two-cluster K-means would mistakenly remove normal data from already-cleaned rock drilling data, the hybrid approach is also tested on borehole A. As shown in Figure 17, the additional use of K-means eliminates three more anomalies, which were not detected by IsoForest, at the cost of losing one normal data point. This indicates that the proposed hybrid method for cleaning soil–rock data does not compromise the data quality of rock-only boreholes, enabling fully automated cleaning of complex datasets in varied geological environments. Based on the above analysis, the efficient real-world implementation of IsoForest or IsoForest+two-cluster K-means is recommended as follows: apply auto-mode IsoForest directly to all boreholes if the site consists solely of rock or has only thin soil layers overlying the rock; if significant soil layers are present, first apply auto-mode IsoForest for initial anomaly detection, followed by two-cluster K-means to eliminate remaining soil drilling data. This strategy avoids the repetition of conducting in rock-only data, and maintains a good balance between true positive rate and false positive rate. It should be noted, however, that the proposed strategy is tailored for rock-dominated drilling conditions. Further research is needed to develop robust cleaning methods for ID data collected in predominantly soil environments.

Table 4 Summary of data cleaning results of statistical methods and machine learning algorithms (borehole B)

| Data cleaning methods | Detected anomalies/total anomalies | Misidentified normal data/total normal data | Detected soil drilling data/total soil drilling data | Optimal hyperparameters |
|---|---|---|---|---|
| 3σ rule | 1/26 | 0/118 | 0/39 | None |
| IQR method | 1/26 | 0/118 | 39/39 | None |
| IsoForest | 22/26 | 2/118 | 5/39 | auto |
| One-class SVM | 23/26 | 16/118 | 17/39 | $nu$=0.3 |
| DBSCAN | 21/26 | 2/118 | 0/39 | $eps$=0.35, $min\_samples$=30 |
| IsoForest combined with two-cluster K-means | 22/26 | 2/118 | 39/39 | None |

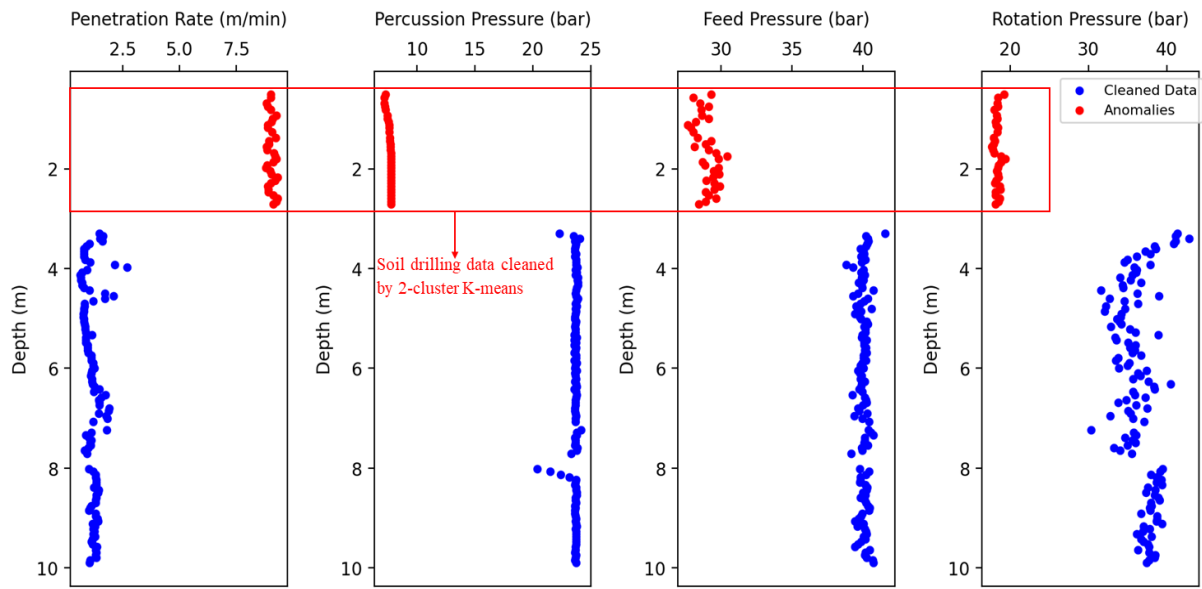

Figure 16 Further moving of the soil drilling data using two-cluster K-means

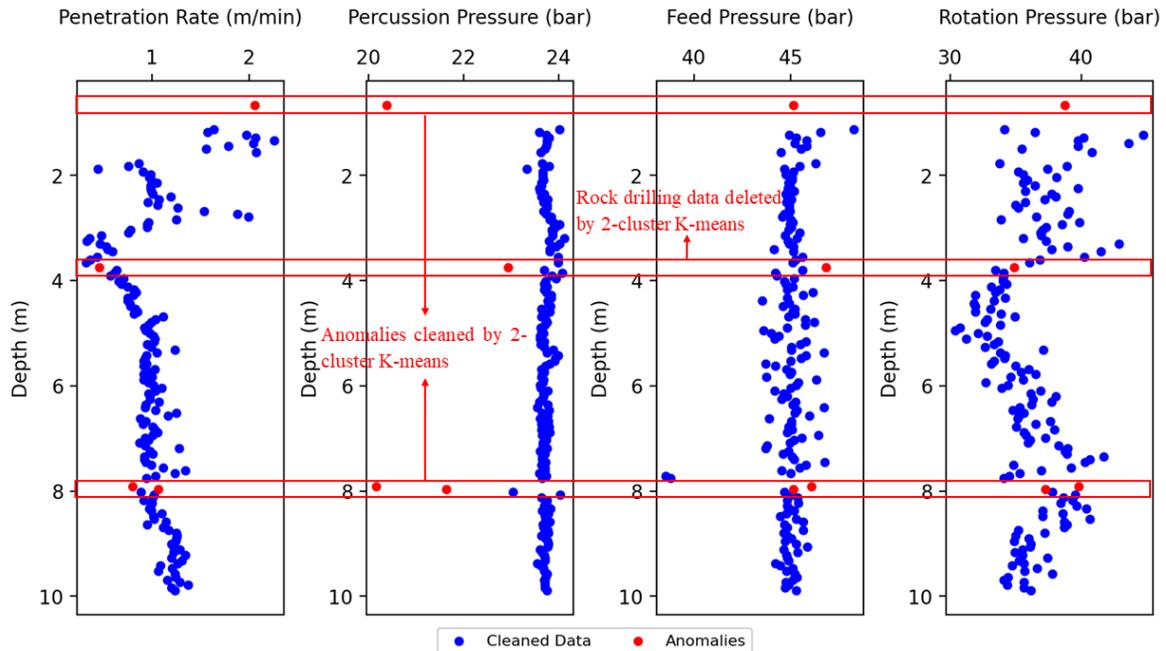

Figure 17 Testing the over-cleaning effect of two-cluster K-means on the cleaned rock drilling data

## 5. Conclusions

In this paper, rock drilling data of ID were analyzed and four major drilling actions that cause anomalies were identified. Five data cleaning approaches, including two statistical methods, the 3σ rule and IQR method, and three machine learning algorithms, including IsoForest, one-class SVM and DBSCAN, were investigated for their cleaning performance on rock drilling data of ID. Two data cleaning tasks were considered, removing anomalies in rock drilling data and removing both anomalies and soil drilling data in mixed rock drilling data. With properly adjusted hyperparameters, machine learning algorithms were proven to outperform statistical methods, among which IsoForest achieved the best balance between removing anomalies and maintaining normal data under the auto mode. For cleaning both anomalies and soil drilling data, a hybrid strategy, which combined IsoForest with two-cluster K-means, was proposed for the more complex data cleaning task. The results showed that the proposed strategy effectively removed both anomalies and soil drilling data without tuning any hyperparameters. The hybrid approach utilized the anomaly detection efficacy of IsoForest and the simplicity of K-means clustering, expanding its application from cleaning pure rock drilling data to more complicated scenarios. The automatic data cleaning approach proposed in this paper can be used to construct large-scale, multi-borehole datasets in the future, facilitating machine learning studies on the complex relationships between drilling data and rock properties.